# Reinterpretation of Lorentz Transformation and resolution of Special Relativity's paradoxes.

Rodrigo de Abreu
Centro de Electrodinâmica e Departamento de Física do IST

## Abstract

Lorentz Transformation is reinterpreted. It is shown that by admitting the existence of a frame of reference with synchronized clocks, we conclude that any other frame of reference that moves related to the first has desynchronized clocks. From this conclusion we will arrive at a new expression to relate the time of different frames of reference. We will also arrive at a new expression to relate the contraction of different frames. We will show that if the maximum speed on the frame with synchronized clocks is the speed of light, then the speed of light varies accordingly to the velocity of the frame of reference. The new interpretation of Lorentz Transformation explains and solves Relativity's paradoxes.

## Synopses

The existence of a frame of reference with truly synchronized clocks, which we will call the resting frame**,** will be postulated. The existence of a maximum speed as measured by the resting clocks will also be assumed. We will show that from this premises and using Lorentz Transformation, the following conclusions can be drawn:

1. If we consider a second frame of reference moving relative to the resting frame then the clocks of that frame are not truly synchronized. Unlike Relativity claims, only one frame of reference can then be considered to be the one at rest and have synchronized clocks.

2. The value by which the clocks are desynchronized varies accordingly to the velocity relative to rest of the moving system of coordinates - the faster the system moves, the larger is the value by which they are desynchronized.

3. Within this frame, Lorentz Transformation gains a new physical meaning that implies, among other things, that:
	- Relations between moving systems of coordinates can only be determined with precision if we know the relation between their absolute velocities (their velocities relative to rest)**.** Lorentz transformation, when directly applied to two moving systems of coordinates, has no physical meaning unless one of the systems has a small absolute velocity.
	- If we have two identical bars and if one is moving faster then the other, then the faster bar will look shorter to the eyes of the slower one, while the slower bar will seem longer to the faster one. If we replace bars by human twins, the same relation will occur for age – one twin will see the other age faster. There is no reciprocity between systems.

   - When the same identical bars, or human twins, are moving away from each other with symmetrical velocities related to rest, they notice absolutely no change when looking at each other. Exactly as when they are stopped related to each other.

4. Relativity's postulate - that a ray of light propagates with the same speed when measured from all systems of coordinates - apparently holds, but only because of the very definition of speed that together with the definition of synchronization becomes tautological.

5. Relativity's paradoxes are the result of trying to interpret Lorentz Transformation without taking in consideration that clocks on moving systems are not synchronized. If this desynchronization is considered when interpreting Lorentz Transformation, then the paradoxes cease to exist.

## 1. Lorentz Transformation implies absolute motion

Relativity allows us to admit any system of coordinates as being at rest and having truly synchronized clocks (*Einstein, A.* [1] p. 125-127). Let us then admit such a system – we'll call it *S*. Let's also admit a second system of coordinates moving with velocity *v* related to *S* – we'll call it *S'*. *S'* only moves right along the *x*-axis of *S* ($y=y'$ and $z=z'$ for every *t*). Each of this systems has an infinite number of clocks. There is a different clock situated on every position of *x*. There's also a different clock for every *x'*. Obviously all the clocks of *S* are synchronized in relation to each other since we began by admitting it. Let us then find out if the same is true for the clocks on *S'*.

Let's analyze a "snapshot" of our world. Let's suppose, for simplicity's sake, that when the clocks on *S* mark the instant *0* (*t=0*), the clock at the origin of *S'* also marks the instant *t'= 0*. Let's also assume that when *t=0* the origin of *S'* is passing on the origin of *S* – this is the "snapshot" we'll analyze. Lorentz Transformation, for this situation, is the following system of equations (where *k* is the maximum speed):

$$x = \frac{x' + vt'}{\sqrt{1 - \frac{v^2}{k^2}}} \qquad (1)$$

$$y = y'$$
$$z = z'$$

$$t = \frac{t' + \frac{v}{K^2} x'}{\sqrt{1 - \frac{v^2}{k^2}}} \qquad (2)$$

Within Relativity's framework, this system allows us to relate any two frames of reference. We'll then use it to relate $S$ with $S'$. Since we know that $t=0$ let us find out $t'$ for any given $x'$, that is, let's find the instant marked by any given clock of $S'$. By making $t=0$, from (2) we get:

$$0 = \frac{t' + \frac{v}{k^2}x'}{\sqrt{1 - \frac{v^2}{k^2}}} \Leftrightarrow t' = -\frac{v}{k^2}x' \qquad (3)$$

We conclude from (3) that $x' \neq 0$ implies $t' \neq 0$, that is, every clock placed away from the origin of $S'$ marks an instant that differs from $0$. Our premise was that the clock on the origin of $S'$ marks the instant $0$, so we conclude that all the other clocks are desynchronized. Since $t'$ is a function of $x'$, all the clocks of $S'$ mark different times between each other. The further a clock is from the one placed on the origin, the larger is the desynchronization between them. We can also see from (3) that the desynchronization of two given clocks is a function of the velocity of the system[1]. The faster a system is moving, the larger is the desynchronization between its clocks[2].

We began by admitting a frame of reference with synchronized clocks. We called this frame the resting frame**.** We concluded that the use of Lorentz Transformation to relate that frame of reference with any moving frame, implies that the clocks on the moving frame are not synchronized. Only one frame of reference has synchronized clocks. That frame of reference is different from all the others. This implies we cannot, alternatively, consider the moving frame to be resting and assume the resting one is moving. One and only one frame of reference can be at rest [2]. On the other side, the value of the desynchronization of a moving frame depends on the velocity relative to the resting one. We conclude then, within Relativity's own framework, that rest and motion are absolute.

## 2. Reinterpreting Lorentz Transformation

Once we understand clocks on moving frames are desynchronized, we can no longer continue to interpret Lorentz Transformation the same way Relativity does. On the other hand, acknowledging this desynchronization allows us to understand the origin of Relativity's Paradoxes. In order to do that, we'll continue to analyze $S$ and $S'$ as we've defined them before.

---

[1] This function of velocity should not be confused with the one that relates the rhythm of clocks between systems of coordinates. Acknowledging the difference between rhythm and synchronism is essential to understanding the problem - notice that two clocks can be moving at the same rhythm without being synchronized.

[2] We've analyzed a specific "snapshot" of our world. It can be easily shown that any other values of $t$ and $t'$ lead us to the same conclusions.

## 2.1 Time between different frames - Solving the Twin Paradox

Accordingly to (3), the instants marked by each clock of S and S', when *t=0*, can be illustrated by Fig.1:

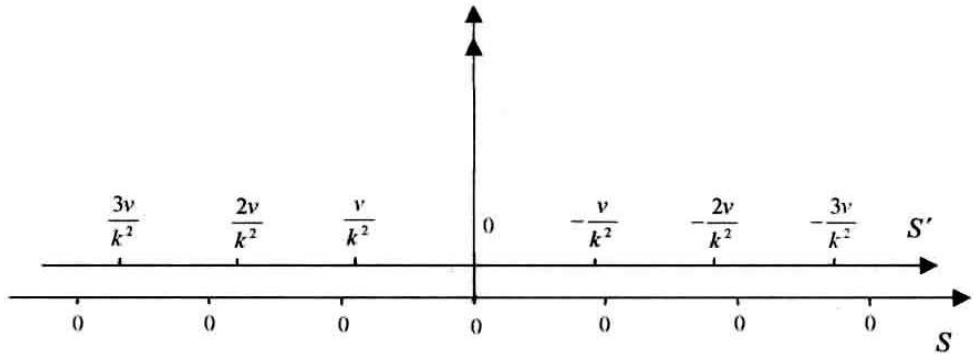

**Fig. 1** This is the snapshot of both frames at *t=0*. Notice that only one of the clocks of S' reads the instant *t'=0*. (Only the clocks of x=0,±1, ±2, ±3 and x'=0,±1, ±2, ±3 are shown)

Let's consider one of the clocks of S', the one on *x'=0* for simplicity. When our clock marks *t'=0*, the clock that corresponds it on S also marks the instant *t=0*. We know S' is moving related to the resting system of coordinates with velocity *v*. As time passes and our clock moves along S, it'll find the clocks of S. We know the clocks of S are synchronized between each other. All the clocks of S mark the same instant at the same moment. This means that by comparing our clock with the clocks it'll find on S as it moves and times goes by, we can conclude which is working faster. That's what we'll do by making *x'=0* on (2). For a given *t'* of our clock, the corresponding clock of S marks the following instant:

$$t = \frac{t'}{\sqrt{1 - \frac{v^2}{k^2}}} \qquad (4)$$

We see that as our clock moves along *S*, it will find the clocks of *S* marking instants that keep getting larger then the one our clock marks. The clock of S' is also working, its just that the clocks of *S* are working faster. Their rhythms are different. The clocks on *S* are getting older faster then those of S' [*Einstein, A.* [1] p.139).

Let's now look at the same situation from the symmetrical point of view. We'll now focus our attention on a single clock of *S* - as before our clock will be the one on the origin of S, *x=0*. When our clock marks the instant *t=0*, the clock that corresponds it on S' is situated on *x'=0* and it also marks *t'=0*. As we had assumed, S and S' are moving related to each other. S is the resting frame, but it's moving related to S'. As it does, our clock will begin to find the clocks of S'. As we've seen before, the clocks of S' are moving slower then those of *S*. We would then expect our clock to find the clocks of S' marking smaller instants, accordingly to the difference in rhythms. But the clocks of S' are not synchronized! They were already marking larger instants when our clock left

*x'=0*. They're just all working at the same rhythm. For a given *t* of our clock (*x=0*), from (1) and (2) the corresponding clock of *S'* marks the following instant

$$t' = \frac{t}{\sqrt{1 - \frac{v^2}{k^2}}} \qquad (5)$$

As we first look at (5), a full symmetry apparently exists - as the clock of *S* moves along *S'*, it also finds clocks marking instants that keep getting larger. But now, it's not because the clocks of *S'* are working faster. It's the desynchronization between the clocks of *S'*, combined with the difference in rhythms, that makes it seem that way. The value of *t'* on (5) is <u>not</u> the time that passed on *S'* when *t* time has passed on our clock. *t'* is solely the instant the clock of *S'* reads when our clock reaches it. We cannot conclude that *S'* is getting older faster then *S*[3]. We must find another way to compare the rhythm of the clock of S with that of the clocks S'.

Since all the clocks of S' are moving at the same rhythm, we can use a single clock of S' for comparison. We can find the time passed on that clock, for a given time passed on the clock of S.

From (1) and (2) we obtain:

$$x' = \frac{x - vt}{\sqrt{1 - \frac{v^2}{k^2}}} \qquad (6)$$

$$t' = \frac{t - \frac{v}{k^2} x}{\sqrt{1 - \frac{v^2}{k^2}}} \qquad (7)$$

Let us go back to *t=0*, when the origins of *S* and *S'* coincide (Fig 1). After a given *t* the origin of *S* arrives at a certain clock of *S'*. Or, in other words, after a given *t*, a clock of *S'* arrives at the origin of *S*. Let's analyze this clock. Since *S'* is moving with velocity *v* related to *S'*, *S* is moving with velocity *–v* related to *S'*. This means that the clock that will arrive, at time *t*, at the origin of *S* is at *x=- v t* when *t=0*. We can calculate the instant $t_1'$ this clock reads when it is still at that position. From (7), by making $t=t_1=0$ and $x=x_1=-vt$ we get,

$$t_1' = \frac{t_1 - \frac{v}{K^2}x_1}{\sqrt{1-\frac{v^2}{K^2}}} = \frac{\frac{v}{K^2}vt}{\sqrt{1-\frac{v^2}{K^2}}} \qquad (8)$$

---

[3] To better visualize the absurd of such a conclusion we can imagine a flight from London to New York on the Concord. Supposing the plane left London at 10:00 local time and arrived at JFK's airport five hours later, the clocks on the airport would still read 10:00 since NY is in a different meridian. But what sense would it make for a passenger to conclude time had stood still since the plane took of?

The instant $t_2'$ the same clock reads when it arrives at the origin of S is given by (5),

$$t_2' = \frac{t}{\sqrt{1-\frac{v^2}{k^2}}} \qquad (9)$$

The difference between these two instants – (8) and (9) - is the time passed on the clock of S',

$$t_2' - t_1' = \frac{t(1-\frac{v^2}{k^2})}{\sqrt{1-\frac{v^2}{k^2}}} = t\sqrt{1-\frac{v^2}{k^2}} \qquad (10)$$

If instead of writing $t_2'-t_1'$ on (10) we simply write $t'$, then (10) is then the same as,

$$t = \frac{t'}{\sqrt{1-\frac{v^2}{k^2}}} \qquad (11)$$

(11) gives us the real time that passed on the clock of S, when a given time $t'$ passed on a clock of S'. (11) is different from (5). $t'$ has a different meaning in each of this expressions. While in (11) $t'$ is the time that passed on S', in (5) it is not. As expected, (11) is the same as (4) - the relation between the rhythms of S and S' doesn't dependent on which frame of reference we place ourselves. We confirm from (11) that the time interval set by a single clock of S' is smaller then that of a clock of S. Time goes by slower for S'. S is getting older faster then S' independently of the point of view of the observer.

It's now clear that the Twin Paradox is the result of a wrong interpretation of Lorentz Transformation. It's a consequence of trying to measure time with desynchronized clocks. While (2) and (7) are symmetrical (as are (1) and (8)), their physical meaning is not the same.

In order to establish an exact relation between the rhythms of two moving frames of reference, we have to know how their clocks are desynchronized. This implies knowing their velocities relative to rest. We can't otherwise be certain which twin will age faster. How can we then establish such a relation?

Let's consider a third system of coordinates, $S''$, with absolute velocity $v_2$. We want to relate $S'$ with $S''$. We already know how to establish the relation between the rhythms of each of the moving frames and the resting one. We know from (4) that:

$$dt' = dt\sqrt{1 - \frac{v_1^2}{k^2}} \qquad (12)$$

and

$$dt'' = dt\sqrt{1 - \frac{v_2^2}{k^2}} \qquad (13)$$

From (12) and (13), we get

$$dt' = dt'' \frac{\sqrt{1 - \frac{v_1^2}{k^2}}}{\sqrt{1 - \frac{v_2^2}{k^2}}} \qquad (14)$$

(14) is the relation between the rhythms of any two frames with absolute velocities $v_1$ and $v_2$. By making $v_2=0$ we obtain the relation between a moving frame and the resting one, which we already knew to be (12). Now, supposing the two systems of coordinates were moving away from each other with identical absolute velocities, then $v_1 = v_2$ and from (14) we'd get:

$$dt' = dt'' \qquad (15)$$

This means that when two twins are moving away from each other with identical absolute velocities, they're aging at exactly the same rhythm. As if they were stopped in relation to each other. If one of this twins were to catch up with the other, he would have to turn around and speed up. For a period of time his absolute velocity would have to be larger then that of the other twin. During that period he would be aging slower then his brother, so when he finally did reach him, he would look younger.

In conclusion, the expression (12) or (13), which within Relativity's framework is used to relate any two frames of reference, doesn't express any physical meaning unless one of the frames has a very small absolute velocity. In order to know the exact relation between the rhythms of the clocks of two moving frames, (14) has to be used. This shows the rhythm of a clock depends on it's absolute velocity.

**2.2 Lorentz Contraction**

The problem with Lorentz Contraction is analogous of that of measuring time between frames. In order to measure the size of a moving material bar on a resting system of coordinates we need to localize both ends of the bar at exactly the same moment (*Einstein, A.* [1] p.128). Once we know those two positions, we can determine the distance between them and we'll get the size of the bar. But since the clocks of a moving system are desynchronized, how can we be sure we're looking at both ends of a bar at exactly the same moment?

Let's consider a bar that when resting related to any system of coordinates has size *L* measured on the *x*-axis of that system. Let's suppose the bar is moving with *S'* that has a speed $v_1$ and that the left end of the bar, *A'*, is fixed on $x_A'=0$. Since the bar is resting related to *S'*, it's right end, *B'*, is on *x'=L*. We want to know the size of the bar as measured in *S*. Since the clocks of *S'* are desynchronized, we must begin by finding the instant marked by the clock situated on *x'=L* when *t=0* (Fig 1). From (3) we get,

$$t' = -\frac{v_1}{k^2} L \qquad (16)$$

Knowing *t'*, from (1) we get the position of *B'* on *S*

$$x = \frac{L + v_1(-\frac{v_1}{k^2}L)}{\sqrt{1-\frac{v_1^2}{k^2}}} = L\sqrt{1-\frac{v_1^2}{k^2}} \qquad (17)$$

Since *A'* is on *x=0*, (17) is exactly the size of the bar measured on *S*. *x* is smaller then *L* and is a function of $v_1$. The faster the bar is moving, the more it gets contracted. Now let's suppose there's an identical bar fixed on *S*. How can we measure it's size from the point of view of *S'* ?

We'll also suppose *A* is on *x=0* and therefore *B* is on $x_B=L$. *S* is moving related to *S'* so, as before, the only way to determine the size of the bar on the *S* frame is to localize its ends at exactly the same time. We'll again suppose that when the left end of the bar, *A*, is on *x'=0*, the clock that's situated on the origin of *S'* marks *t'=0*. We want to know where *B* is at the same moment. But now we arrive at the problem we've come upon before - the clocks of *S'* are desynchronized. When the clock of *x'=0* marks *t'=0*, the clock of *S'* situated next to *B*, marks a different instant. If we where to try to find *B* for *t'=0*, we wouldn't be localizing it for the same moment in time. We would be localizing it for a later moment and so the bar would've already moved. We would find an unreal smaller value for it's size. One way to be sure we're localizing *B* at exactly the same moment is to use the clocks of the resting frame. Only that frame of reference has synchronized clocks. Since when the clock of the origin of *S'* marks *t'=0* the clocks of *S* also mark *t=0*, we'll localize *B'* for *t=0*. From (6) by making *x=L* and *t=0* we get:

$$x' = \frac{L}{\sqrt{1 - \frac{v_1^2}{k^2}}} \qquad (18)$$

Has expected the bar looks larger from the point of view of *S'*. The faster *S'* is moving the larger the bar of *S* will seem to be. This only seems natural since we had already seen *S'* is contracted because of it's velocity relative to *S*.

It becomes clear that, like time, the contraction of a bar can only be determined using the frame with synchronized clocks as reference - the resting frame. How can we then determine the size of a moving bar on an also moving frame of reference?

Let's suppose we want to measure the size of the bar of *S'* on a system of coordinates *S''* moving with velocity $v_2$. We must begin by localizing *B'* on *S*, which we already know to be (17). Now all we need to do is localize the obtained *x* on *S''*, at the same moment *t=0*. Supposing *A'* is on *x''=0*, from (6) (the *x'* of (6) is now *x''*) substituting *x* for (17) and making *t=0* we get:

$$x'' = \frac{L\sqrt{1 - \frac{v_1^2}{k^2}}}{\sqrt{1 - \frac{v_2^2}{k^2}}} \qquad (19)$$

(19) is the size of the bar measured on any system of coordinates with velocity $v_2$, when the bar is moving with velocity $v_1$. By making $v_2=0$ we get Lorentz Contraction, which means we're measuring the bar on the resting frame. When $v_1=0$ the bar is fixed on the resting system. By making $v_1 = -v_2$ or $v_1 = v_2$ the bar and the frame we're measuring it in are traveling with the same speed related to rest – they are equally contracted and so the size of the bar is *L*.

What happens when we localize *B'* on *S''* for *t''=0?* Since the clocks of *S''* are also desynchronized we are localizing it for a different moment in time. For a given *t*, the position of *B'* on *S* is:

$$x = L\sqrt{1 - \frac{v_1^2}{k^2}} + v_1 t \qquad (20)$$

On the other hand, from (7) (the *t´* of (7) is now *t´´*), when *t´´=0*, *t* is given by

$$0 = \frac{t - \frac{v_2}{k^2}x}{\sqrt{1 - \frac{v_2^2}{k^2}}} \qquad (21)$$

By substituting *x* from (20) we get

$$0 = \frac{t - \frac{v_2}{k^2}L\sqrt{1 - \frac{v_1^2}{k^2}} - \frac{v_2 v_1}{k^2}t}{\sqrt{1 - \frac{v_2^2}{k^2}}} \Leftrightarrow t = \frac{\frac{v_2}{k^2}L\sqrt{1 - \frac{v_1^2}{k^2}}}{1 - \frac{v_1 v_2}{k^2}} \qquad (22)$$

We can now substitute *t* for (22) on (20),

$$x = L\sqrt{1 - \frac{v_1^2}{k^2}} + v_1 \frac{\frac{v_2}{k^2}L\sqrt{1 - \frac{v_1^2}{k^2}}}{1 - \frac{v_1 v_2}{k^2}} = \frac{L\sqrt{1 - \frac{v_1^2}{k^2}}}{1 - \frac{v_1 v_2}{k^2}} \qquad (23)$$

Now we know *t* and *x*. Substituting on Lorentz Transformation (6) (the *x´* of (6) is now *x´´*), we find *x´´*

$$x´´= \frac{\frac{L\sqrt{1 - \frac{v_1^2}{k^2}}}{1 - \frac{v_1 v_2}{k^2}} - \frac{v_2^2}{k^2}\frac{L\sqrt{1 - \frac{v_1^2}{k^2}}}{1 - \frac{v_1 v_2}{k^2}}}{\sqrt{1 - \frac{v_2^2}{k^2}}} = \frac{L\sqrt{1 - \frac{v_1^2}{k^2}}\sqrt{1 - \frac{v_2^2}{k^2}}}{1 - \frac{v_1 v_2}{k^2}} \qquad (24)$$

(24) can been written in the form

$$x´´= L\sqrt{1 - \frac{v_R^2}{k^2}} \qquad (25)$$

where

$$v_R = \frac{v_2 - v_1}{1 - \frac{v_2 v_1}{k^2}} \qquad (26)$$

is the relative speed between $S'$ and $S''$.

As $B$ moves along $S''$ it finds clocks marking different values. (25) is the position of the clock that marks $t''=0$ when $B$ reaches it. Naturally by that time $A$ has also moved. (25) has the form of Lorentz Contraction, but as we've seen, its physical meaning is different - (25) is not the size of the bar, it's a position of the bar's right end after it has already moved.

In conclusion Lorentz Contraction can only measure the size of a bar on the resting frame of reference. When Lorentz Contraction is established from any other frame of reference it's physical meaning is not the same - since the clocks of any moving frame of reference are not synchronized, the ends of the bar are being localized at two different moments in time - we're not finding the bar's true size. Only when that frame of reference has a very small absolute velocity do those values become similar. We've also seen that unlike Relativity claims, identical bars don't see each other as looking smaller (*Einstein, A.* [1] p. 138). The faster bar sees the other larger and the slower one sees the other smaller. On the other hand, we find that when two bars are moving with symmetrical absolute velocities they notice absolutely no contraction between each other.

## 3. The inconstancy of the speed of light - reinterpreting Einstein's definitions

Let's begin by assuming (*Einstein, A.* [1] p.134) the maximum possible speed k on the frame of reference we considered to be at rest is the speed of light ($k=c$). This means that if a ray of light where to leave the origin of $S$ when $t=0$, it would reach $x=L$ at $t=L/k$. All the clocks of $S$ would be marking that same instant by that moment.

How can we know the speed of that same ray of light related to a moving frame of reference $S'$? Through (20) and $x=kt$ and using Lorentz transformation it can be easily shown that the instant $t'$ marked by the clock of $S'$ situated on $x'=L$ would also be $L/k$ when the ray of light reached it. This is true for any velocity of $S'$. This could lead us to think the speed of light is the same for every frame of reference. But we can't forget the clocks of every moving frame of reference are desynchronized. When the ray of light left $x'=0$ the clock situated on $x'=0$ marked $t'=0$, but the one on $x'=L$ was marking a smaller value. From (3) it marked:

$$-\frac{v}{k^2} L \qquad (27)$$

This means that *(L/k-0)* is not the time the ray of light takes to reach $x'=L$. The time it takes for the light to reach that point can be calculated by only looking at the clock of

*x'=L*. It is the difference between its instant when the light reaches it and the instant it marked when the light left *x'=0*, (27). This is what we get:

$$\frac{L}{k} - (-\frac{v}{k^2}L) = \frac{L}{k}(1 + \frac{v}{k}) \qquad (28)$$

The maximum speed on *S'* is then *L* divided by that time,

$$\frac{L}{\frac{L}{k}(1+\frac{v}{k})} = \frac{k}{1+\frac{v}{k}} = \frac{k(1-\frac{v}{k})}{(1+\frac{v}{k})(1-\frac{v}{k})} = \frac{k-v}{1-\frac{v^2}{k^2}} \qquad (29)$$

(29) is the maximum speed measured on any frame of reference with velocity *v* related to rest. The maximum speed depends on the velocity of the frame. If we consider that the maximum speed is the speed of light, then the speed of light changes accordingly to the speed of the frame of reference and its direction. It's value is *k* only on the resting frame. When the ray of light is traveling in the same direction of the frame we're measuring it in, its speed varies between *k* and *k/2*. As assumed, on the resting frame the speed of light is *k*. As the speed of the frame approaches *k*, the speed of the ray of light measured on that frame, becomes *k/2*. When the ray of light is traveling in the opposite direction of the frame of reference, its speed varies between *k* and infinity - as the speed of the frame approaches *k* the speed of that same ray of light tends to infinity. This results leads us to another interesting conclusion. When a ray of light is first emitted with the direction of the moving frame and then gets reflected by a mirror, it travels until it reaches the mirror with one speed and then it travels back in the opposite direction with a different speed. We can determine its average speed since it leaves *x'=0* until it gets back to that point. We'll consider the mirror is on *x'=L* when the light reaches it. We already know both speeds from (29) and the duration of both routes from (28). The average speed, where *T* is the total time is:

$$(\frac{k}{1+\frac{v}{k}} \times \frac{L}{k}(1+\frac{v}{k}) + \frac{k}{1-\frac{v}{k}} \times \frac{L}{k}(1-\frac{v}{k}))/T \qquad (30)$$

*T* is

$$\frac{L}{k}(1+\frac{v}{k}) + \frac{L}{k}(1-\frac{v}{k}) = \frac{2L}{k} \qquad (31)$$

Substituting *T* in (30) we obtain *k*.

This means the average speed of a ray of light that leaves one point and then returns is the same on every frame of reference. It is always *k*. But, has we've seen, this doesn't mean the instantaneous speed of light doesn't vary accordingly to the speed of the frame.

Einstein's synchronization method depends on the emission of a ray of light. Einstein definition [*Einstein, A* [1] p. 125-128] implies that a clock situated at a distance *L* from the point of emission marks the value *L/k* by the time light reaches it. Every clock of any frame of reference obeys this rule. But at the same time, Lorentz Transformation implies the clocks of *x'=0* and *x'=L* are desynchronized between each other and, of course, will continue to be when light reaches *x'=L*. Once we admit the speed of light varies accordingly to the velocity of a frame, the reason for this desynchronization becomes clear[4]. Except for the resting frame, Einstein's synchronization method doesn't truly synchronize clocks. By defining speed as a function of the difference between the instants of two desynchronized clocks, Einstein created a new definition of speed since that difference is not really the time light takes to reach the clock. We can't say this definition is wrong, but we must understand that by using it, we will necessarily conclude the speed of light is the same on every frame of reference[5]. By using the usual definition of speed, we will arrive at different conclusions. Both definitions can be used, but in order to arrive at a physical interpretation, we must know which we're using.

Relativity tells us that if two events occur at the same time on one frame of reference, then they don't from the point of view of a different frame - what is simultaneous on one frame of reference isn't on another *[Einstein, A [1] p 125-130, 138]*. Again, this apparent paradox is the result of not acknowledging the desynchronization. As we've seen, by trying to localize a bar, "the same moment" on a moving frame is defined by clocks marking different instants. On a moving frame, talking about two clocks reading the same instant is talking about two events that don't occur at the same moment – the events are not truly simultaneous. Except for the resting frame, the definition of simultaneity can't depend on clocks reading the same instant.

Essentially the problem with Relativity is one of interpretation. Relativity's postulates are incompatible with Lorentz Transformation. Clocks of a moving frame that were set using the speed of light, are not synchronized between each other. Lorentz transformation implies that. This it self is not a problem. As long as we know precisely how those clocks are desynchronized, we can use them just as well as if they were synchronized. What is a problem is not acknowledging that desynchronization and assuming every frame of reference has synchronized clocks. Lorentz transformation cannot be used under that assumption. This is has been happening within Relativity. Unless we understand that we are dealing with desynchronized clocks and unless we have in mind the meaning of the definitions we're using, we're likely to continue to make predictions about the physical world that will lead us to incomprehensible paradoxes.

---

[4] Accordingly to Einstein's definition of synchronization, when a ray of light reaches a clock at distance *L*, it sets the clock at *t'=L/c*. Of course, if the speed of light is not really c on that frame, then *L/c* is not the time it takes for the light to travel the distance *L*. On the other hand, the clock that was reading *t'=0* when the light was emitted will read the real time the light took to reach *L*. The clocks will be reading different instants.

[5] *"We thus see that the velocity of transmission relative to the reference-body K' is also equal to c. The same result is obtained for rays of light advancing in any other direction whatsoever. Of course this is not surprising, since the equations of the Lorentz transformation were derived conformably to this point of view."* Einstein, Albert. *Relativity: The Special and General Theory.* New York: Henry Holt, 1920; Bartleby.com, 2000. www.bartleby.com/173/. [Date of Printout].